\documentclass[prx,aps,twocolumn, longbibliography]{revtex4-2}
\usepackage{amsmath,amssymb, graphicx}
\usepackage{bm,times, color}
\usepackage[utf8]{inputenc}
\usepackage[colorlinks]{hyperref}
\usepackage[svgnames]{xcolor}
\hypersetup{linkcolor=DarkBlue, citecolor=DarkBlue, filecolor=black, urlcolor=DarkBlue}
\usepackage{comment, mathrsfs,slashed}
\def \p{\partial}
\def \dag{\dagger}
\def \mb{\mathbf}

\def \lan{\langle}
\def \ran{\rangle}

\def \ga{\gamma}

\def \ms{\mathsf}
\begin{document}
\preprint{EFI-xx}
\title{Time-reversal odd transport in bilayer graphene: Hall conductivity and Hall viscosity}

\author{Wei-Han Hsiao}
\affiliation{Kadanoff Center for Theoretical Physics, University of Chicago,
Chicago, Illinois 60637, USA}
\date{February 2021}

\begin{abstract}
We consider the time-reversal odd dynamics of the bilayer graphene at low energies in the quantum Hall regime. A generating functional for the effective action that captures the electromagnetic response to all orders in momentum and frequency is presented and evaluated to the third order in the space-time gradient $\mathcal O(\p^3)$. In addition, we calculate the Hall viscosity and derive an explicit relationship with the $q^2$ coefficient of the Hall conductivity. It is reminiscent of the Hoyos--Son relation in Galilean invariant systems, which can be recovered in the limit of large filling factor $N$.
\end{abstract}

\maketitle

\section{Introduction}
The family of multilayer graphite is one of the most intriguing paradigms in the realm of modern condensed matter. The simplest model in the group, graphene, has drawn enormous attention from both theoretical and experimental communities\cite{RevModPhys.81.109}. Its linear dispersion at low energy makes it a low-dimensional example of a particle-hole symmetric ultrarelativistic Dirac fermion, the unconventional electronic property of which distinguishes it from other semiconductors made up of ordinary nonrelativistic quasiparticles. In particular, when placed in a magnetic field, each Dirac point possesses the anomalous Hall conductivity $\frac{1}{2}\frac{e^2}{h}$ at the filling factor $\nu = 0$ \cite{PhysRevLett.95.146801}, which is uniquely connected with the filled Fermi sea of Dirac particles. 

\begin{figure}
 \includegraphics[width=0.9\linewidth]{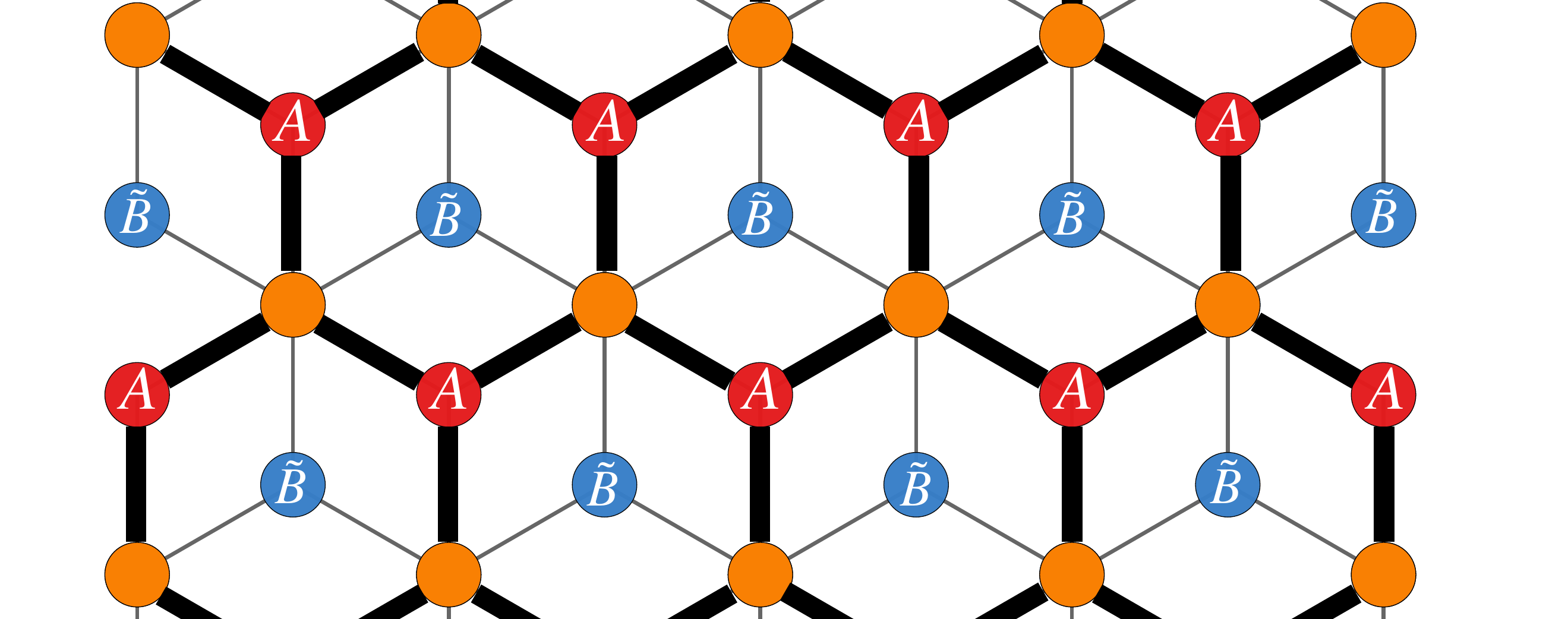}
 \caption{The schematic plot for the $AB$-stacked bilayer graphene. The thick bonds represent the upper lattice, while the thin ones represents the lower layer. The overlapped orange sites have labels $\tilde A = B$. The plot was generated using the PYTHON package PYBINDING \onlinecite{dean_moldovan_2020_4010216}.}
 \label{fig2}
\end{figure}

Even more fruitful physics emerges as one stacks two layers of graphene on top of each other. In the $AB$-stacked configuration as shown in Fig. \ref{fig2}, the low-energy projection of the model forms another family without any relativistic analog: a particle-hole symmetric two-band semiconductor with parabolic dispersion\cite{2013RPPh76e6503M}. It also serves as a model possessing a Fermi surface with a Berry phase of $2\pi$ and is also a candidate for the dual description of the $\nu = 1$ fractional quantum Hall state in the context of fermion-vortex duality \cite{PhysRevLett.117.136802}.  

As the background magnetic field is turned on, Landau levels form in bilayer graphene as well. The spectrum, nevertheless, possesses two zero-energy bands\cite{PhysRevLett.96.086805}. This feature reshapes our understanding of low energy physics in quantum Hall systems \cite{Novoselov2006}. In particular, in a large magnetic field, the conventional wisdom and machinery of the lowest Landau level projection has to be modified since the low-energy sector contains more than holomorphic wave functions in the symmetric gauge and the system can host exotic quantum phases in the lowest Landau level \cite{PhysRevLett.112.046602}. The dielectric property and the low energy excitations in the zero-energy bands under this circumstance were investigated in Ref. \onlinecite{PhysRevB.77.195423} and Refs. \onlinecite{PhysRevB.79.165402}.

This paper concentrates on the time-reversal odd responses of bilayer graphene to dynamical and inhomogeneous external perturbations in a strong out-of-plane magnetic field in (2+1) dimensions, which has been partially addressed recently in an independent work \onlinecite{2019arXiv190909608I}. Here, we adopt a different approach that systematically generates the effective action as a functional of external gauge field $A_{\mu}$ to all orders in momentum and frequency. Using this, we compute the effective action as in the gradient expansion to explore the large scale dynamics. As a result, we show that for the low-energy model of bilayer graphene in a background magnetic field, the Hall conductivity in an inhomogeneous and time-dependent electric field $E_i(\omega, q)$ at filling factor $N$ is 
\begin{align}
\sigma_H(\omega, q)\approx \frac{N}{2\pi} \bigg(1 + (q\ell)^2 \frac{1-3N^2}{4N}+ \frac{\omega^2}{\omega_c^2}\frac{2N^2-1}{2N^2}\bigg),
\end{align}
where $\ell$ and $\omega_c$ are the magnetic length and cyclotron frequency. We also establish a relationship between the Hall viscosity and the coefficient of the $(q\ell)^2$ term. In particular, with a high-energy Landau level cutoff $N_c$, the static ($\omega =0$) result reduces to 
\begin{align}
\label{main2}\sigma_H(q)= \sigma_H(0) + (q\ell)^2[\ell^2\eta_H -\epsilon''(B) - \mathcal O(\ln N_c)].
\end{align} 
It is reminiscent of the Hoyos--Son relation\cite{PhysRevLett.108.066805} for Galilean invariant systems. An explicit formula is derived to compute higher-order corrections in $N^{-1}$ and $N_c^{-1}$. 

For Galilean invariant systems, the Hoyos--Son relation has been systematically and thoroughly discussed. For instance, Ref. \onlinecite{PhysRevB.86.245309} delivers a concrete relationship between the non-local Hall conductivity and the Hall viscosity. This framework has recently been generalized to anisotropic and lattice-regularized systems in Ref. \onlinecite{PhysRevX.10.021005}. As for graphene-like systems, Ref. \onlinecite{PhysRevB.94.125427} tackles a similar problem for a single layer in a strong magnetic field and low temperature, whereas Ref. \onlinecite{PhysRevB.92.115426} reports a Hoyos--Son-like relation in the interaction-dominating regime using the hydrodynamic approach in the absence of a magnetic field. This paper aims to provide a different generalization by considering the zero-temperature dynamics of a rotationally invariant and particle-hole symmetric system breaking both Galilean and Lorentz symmetries. 

This paper is organized as follows: In Sec. \ref{EFT} we first review the methodology of the one-loop effective action from a microscopic model and introduce the low-energy two-band model for bilayer graphene. We then start presenting this work by first deriving the Feynman rules and the generating functional for the polarization tensor. What follows is the time-reversal odd effective action computed to cubic order in space-time gradients with emphasis on the coefficient of Hall conductivity at the order of $(q\ell)^2$ and an investigation of its relationship with the Hall viscosity. In Sec. \ref{VisSus}, we construct the stress tensor and compute the Hall viscosity and orbital magnetic susceptibility for our model. This provides numerical support for the observation established in Sec. \ref{EFT}. Finally we revisit the conductivity tensor using the Kubo formula in Sec. \ref{KuboCurrent} and derive an exact algebraic relation that connects the Hall conductivity and the Hall viscosity in the absence of space-time symmetry. We then conclude the paper. Details of the computations and an alternative derivation of the stress tensor are given in the appendixes for completeness.

\section{the effective action}\label{EFT}
\subsection{Methodology}
The electromagnetic response can be compactly summarized in the effective action as a functional of the $U(1)$ gauge potential. Starting with a microscopic model for the matter field $\psi$ defined by the action $S[\psi]$, one gauges the charge symmetry by coupling the charge and current densities $j^{\mu} = (\rho,j^{i} )$ with the gauge field $A_{\mu}$. The effective action is defined as follows. 
\begin{align}
\label{Seff}\mathcal S_{\rm eff}[A_{\mu}] = -i\ln \int \mathscr D\psi^{\dag}\mathscr D\psi\, e^{iS[\psi]+i\int j_{\mu}A^{\mu}}.
\end{align}
If the external electric and magnetic fields $\mb E$ and $B$ take constant values, Eq.~\eqref{Seff} can be computed and serves as an example of Euler--Heisenberg effective action \cite{KATSNELSON2013160}. As far as the linear response is concerned, $\mathcal S_{\rm eff}$ is usually expanded as a multinomial in $A_{\mu}$. In the $d$-dimensional Fourier space, the effective action under a Gaussian approximation assumes the form
\begin{align}
\mathcal S_{\rm eff} [A_{\mu}]= &\int \frac{d^dq}{(2\pi)^d}\bigg[\bar{j}^{\mu}A_{\mu} + A_{\mu}(-q)\Pi^{\mu\nu}(q)A_{\nu}(q)\bigg],
\end{align}
where $\bar{j}$ denotes the average charge in the ground state. The polarization tensor $\Pi^{\mu\nu}(q)$ encodes the response functions. 

In (2+1) dimensions, the gauge symmetry alone fixes the form of the effective Lagrangian $\mathscr L_{\rm eff}$. Organized by the number of derivatives, the Lagrangian is: 
\begin{align}
\mathscr L_{\rm eff} = &\frac{k}{4\pi} A\, dA\, +\frac{\epsilon}{2}\mb E^2 -\frac{1}{2\mu}B^2 \notag\\
& + \alpha \mb E\cdot(\nabla B)+\beta\epsilon^{ij}E_i\p_tE_j+\mathcal O(\p^4).
\end{align}
Computing the functional determinant for a specific model yields the parameters $k, \epsilon,\mu$, $\alpha$, and $\beta$. 

\begin{figure}
 \includegraphics[width=1.0\linewidth]{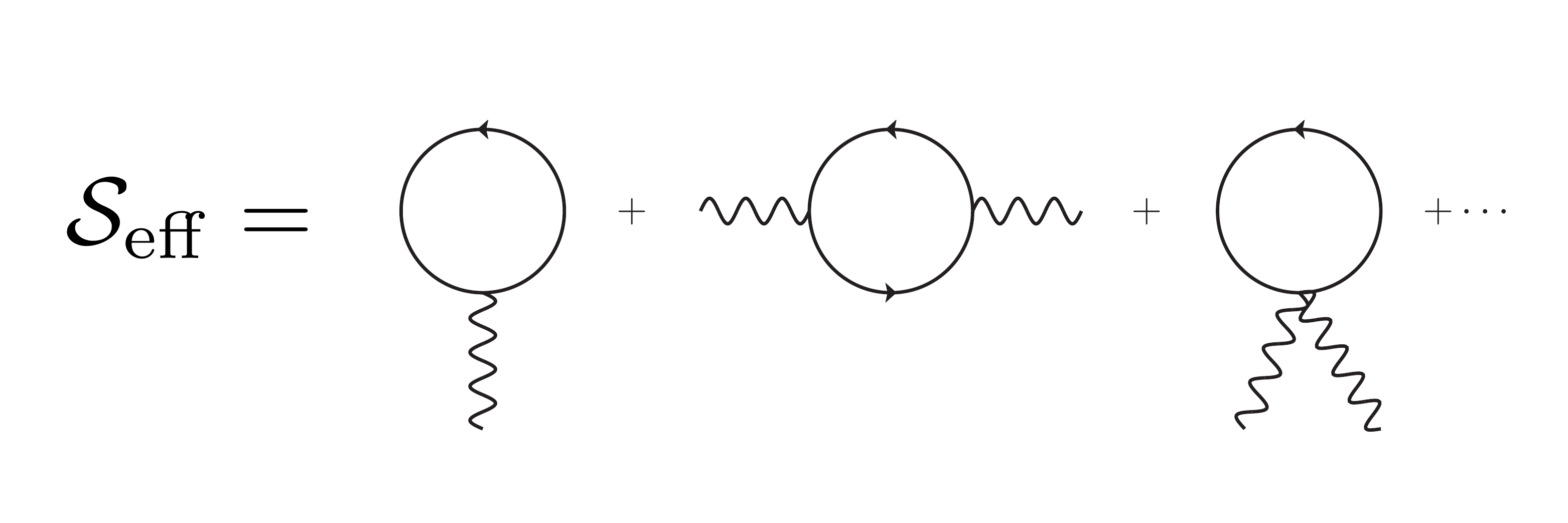}
 \caption{Diagrammatic expansion of the effective action $\mathcal S_{\rm eff}$. Each solid line represents a Feynman propagator $D$ and each wavy line corresponds to the insertion of the external potential $v$. The first and the third diagrams correspond to the trace $\mathrm{tr}[Dv]$, and the diagram in the middle corresponds to the two-point trace $-\frac{i}{2}\mathrm{tr}[DvDv]$.}
 \label{fig1}
\end{figure}

For a fermionic system, computing Eq.~\eqref{Seff} amounts to evaluating the functional determinant of the fermion action. Formally, one can decompose the microscopic Lagrangian for the fermion $\psi$ into the free part $i\psi^{\dag}D^{-1}\psi$ and the potential part $-v(x)\psi^{\dag}\psi$. Such a decomposition is straightforward for a free fermion system in an external potential. If a two-particle interaction is present, one can first introduce an auxiliary field by the Hubbard--Stratonovich transformation to decompose the two-particle term. In this way the fermion part of the theory can be integrated in the path integral, resulting in a functional of $v$. Expanding the functional to the second order of $v$, it reads
\begin{align}
&\mathcal S_{\rm eff} = -i\mathrm{tr}\ln[D^{-1}+iv]\notag\\
\label{LoopExpansion}= & -i\mathrm{tr}\ln [D^{-1}] + \mathrm{tr}[Dv] - \frac{i}{2}\mathrm{tr}[DvDv] + O(v^3).
\end{align}
This formula has an intuitive diagrammatic representation shown in Fig. \ref{fig1}. The inverse of action $D$ corresponds to the Feynman propagator of the free theory and the potential $v$ is the vertex. One can systematically compute Eq.~\eqref{LoopExpansion} using the standard perturbation methods in quantum field theory.
\subsection{The model for bilayer graphene}
Let us now apply the above machinery to the low-energy model of $AB$-stacked bilayer graphene \cite{PhysRevLett.96.086805}. We depict the lattice structure in Fig. \ref{fig2}. The minimal low-energy model for each valley in the Brillouin zone contains two copies of the Dirac fermions and a hopping amplitude $2m_{\star}$ bridging sites $B$ and $\tilde A$. In the low energy regime $\omega\ll m_{\star}$ , the four-band model can be projected to the two dominant bands. 

\begin{figure}
 \includegraphics[width=0.9\linewidth]{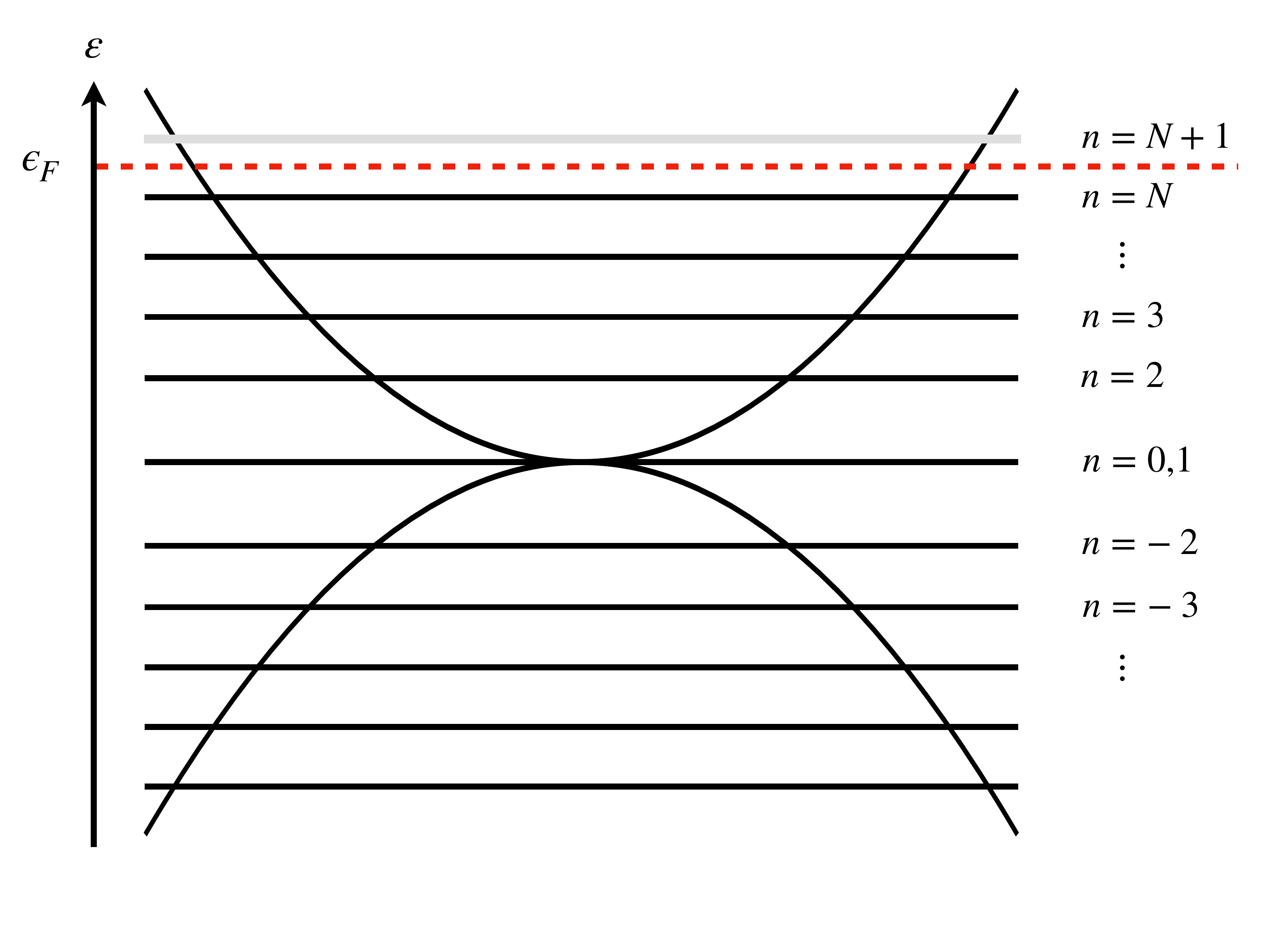}
 \caption{The Landau level spectrum of Eq.~\eqref{spectrum} and the Landau level indices associated with each band. The parabolas depict the dispersion in the absence of the background magnetic field. The red dashed line represents the Fermi energy $\epsilon_F$, separating occupied black bands and the empty gray one.}
 \label{fig3}
\end{figure}

The model for valley $K$ is 
\begin{align}
\label{model}H = -\frac{1}{2m_{\star}}\begin{pmatrix} 0 & \pi^2 \\ (\pi^{\dag})^2 & 0 \end{pmatrix}, \psi_K = \begin{pmatrix}\phi_A\\ \phi_{\tilde B}\end{pmatrix}
\end{align}
where $\pi = \pi_x -i\pi_y$ and $\pi_i = p_i + A_i$ is the kinematic momentum. $\psi_K$ is the spinor storing the dominant two bands. For this model, $k$ and the dielectric constant $\epsilon$ were discovered in Ref. \onlinecite{PhysRevB.77.195423}. The focus of this paper is the computation of $\alpha$ and $\beta$. 
We first derive the propagator $D$ for model~\eqref{model} in a magnetic field. Turning on a finite background magnetic field in the symmetric gauge $(\bar A_x, \bar A_y) = \frac{B}{2}(-y,x)$, the spectrum of the system consists of non-trivial Landau levels. It also introduces the length scale of magnetic length $\ell = B^{-1/2}$. To solve the spectrum, it is convenient to introduce the ladder operators
\begin{subequations}
\begin{align}
& a= \frac{i\ell}{\sqrt 2}\pi\\
& a^{\dag} = -\frac{i\ell}{\sqrt 2}\pi^{\dag},
\end{align}
\end{subequations} 
which satisfies $[a,a^{\dag}] = 1$. The Hilbert space of the $(\widehat{\mb r}, \widehat{\mb p})$ operators is then organized partly using the eigenstates of the operator $\widehat n = a^{\dag}a$, $\{|n\ran| n\in \mathbb Z^+\}$. The Hamiltonian~\eqref{model} can then be shown to have the eigenstates \footnote{We use the notation $|\cdot)$ to denote the doublet wave functions.}
\begin{subequations}
\begin{align}
\label{zerobands}& |0) = \begin{pmatrix} 0 \\ |0\ran \end{pmatrix}, |1) = \begin{pmatrix} 0 \\ |1\ran \end{pmatrix}\\
\label{otherbands}&|n)=\frac{1}{\sqrt{2}}\begin{pmatrix} \mathrm{sgn}(n) ||n|-2\ran \\ ||n|\ran \end{pmatrix}, |n|\geq2,
\end{align}
\end{subequations}
associated with the spectrum (Fig. \ref{fig3})
\begin{align}
\label{spectrum}\varepsilon_n  =\mathrm{sgn}(n) \omega_c\sqrt{|n|(|n|-1)}
\end{align} 
with the cyclotron frequency $\omega_c = B/m_{\star}$.
The degeneracy of each Landau level, as in a two-dimensional electron gas (2DEG) in the symmetric gauge, is encoded by the other set of ladder operators $[b,b^{\dag}] = 1$ which generates the angular momentum $\widehat{\ell}_z = \frac{1}{2}(\widehat{\mb r}\times\widehat{\mb p}-\widehat{\mb p}\times\hat{\mb r})_z$. (See Ref. \onlinecite{jain_2007}.) Note that in this system we no longer treat $\widehat{\ell}_z$ as the canonical angular momentum. The algebraic structure of the second pair of commuting ladder operators still holds. Denoting the Hilbert space with $\frac{1}{\sqrt{m!}}(b^{\dag})^m|0\ran = |m\ran$, the complete basis is $\{|n)|m\ran\equiv|nm)\}.$ Writing $\xi_n = \varepsilon_n-\mu$, the inverse of the kernel $iD^{-1} = i\p_t - \sum_{n,n}(\varepsilon_{n}-\mu)|nm)(nm|$ is
\begin{align}
D(t,t') = \int \frac{d\Omega}{2\pi}e^{-i\Omega(t-t')}\sum_{n,m}\frac{i|nm)(nm|}{\Omega-\xi_n + i\epsilon\, \mathrm{sgn}(\xi_n)}.
\end{align}
Next, we turn on the perturbation on top of $\bar A_{\mu} \to \bar{A}_{\mu}+A_{\mu}$, leading to the variation of the Hamiltonian $H\to H + v(t,\mb x)$, where the vertex $v$ is 
\begin{align}
v(t,\mb x) =&  A_0 -\frac{1}{2m_{\star}}\{\widehat{\Pi}_i,A_i\} & \notag\\
&- \frac{1}{2m_{\star}}\begin{bmatrix} 0 & (A_x-iA_y)^2 \\ (A_x+iA_y)^2 & 0 \end{bmatrix},
\end{align}
with the momentum operators
\begin{subequations}
\begin{align}
& \widehat{\Pi}^x = \begin{pmatrix} 0 & \pi \\ \pi^{\dag} & 0 \end{pmatrix},\\
& \widehat{\Pi}^y = \begin{pmatrix} 0 & -i\pi \\ i\pi^{\dag} & 0 \end{pmatrix}.
\end{align}
\end{subequations}
In order to perform a gradient expansion, the Fourier transform needs to be introduced properly, since the coordinates $\mb x = (x,y)$ are treated as operators on Hilbert space. Given a $c$-valued vector $\mb k$, we recognize that 
\begin{align}
e^{i\mb k\cdot\mb x} = e^{-|\ms k|^2}e^{ia^{\dag}\ms k}e^{ia\bar{\ms k}}e^{ib^{\dag}\bar{\ms k}}e^{ib\ms k},
\end{align}
where the dimensionless complex momenta are $\ms k = \frac{\ell}{\sqrt 2}(k_x-ik_y)$ and $\bar{\ms k} = \ms k^*$. Each operator in $v(t,\mb x)$ is Fourier expanded as an integral of these exponential operators weighted by the Fourier coefficients. For concrete instances, we Fourier transform terms which are linear in the external field. 
\begin{widetext}
\begin{subequations}
\begin{align}
\label{v0} A_0 = &  \int \frac{d^3k}{(2\pi)^3}e^{-i\omega t-|\ms k|^2}\widetilde{A}_0(k) e^{ia^{\dag}\ms k}e^{ia\bar{\ms k}}e^{ib^{\dag}\bar{\ms k}}e^{ib\ms k},\\
\label{v1} \frac{-1}{2m_{\star}}\{\widehat{\Pi}_x, A_x\} = &\int \frac{d^3k}{(2\pi)^3}\frac{ie^{-|\ms k|^2-i\omega t}}{\sqrt 2m_{\star}\ell}\widetilde{A}_x(k)e^{ib^{\dag}\bar{\ms k}}e^{ib\ms k} \{ \begin{pmatrix} 0 & a\\ -a^{\dag} & 0\end{pmatrix}, e^{ia^{\dag}\ms k}e^{ia\bar{\ms k}}\},\\
\label{v2} \frac{-1}{2m_{\star}}\{\widehat{\Pi}_y, A_y\} = &\int \frac{d^3k}{(2\pi)^3}\frac{e^{-|\ms k|^2-i\omega t}}{\sqrt 2m_{\star}\ell}\widetilde{A}_y(k)e^{ib^{\dag}\bar{\ms k}}e^{ib\ms k} \{ \begin{pmatrix} 0 & a\\ a^{\dag} & 0\end{pmatrix}, e^{ia^{\dag}\ms k}e^{ia\bar{\ms k}}\}.
\end{align}
\end{subequations}
Because of the $\pm i$ involved in the definitions of $(a, a^{\dag})$, perturbation in the $x$ direction now resembles $\sigma_y$ and vice versa. This twist will lead to an extra minus in the effective action. We have built the machinery for computing the traces in Eq.~\eqref{LoopExpansion}. As far as the response functions are concerned, the second diagram in Fig. \ref{fig1} is the only nontrivial one. The first diagram gives rise to the ground-state charge density, while in the third diagram, the contact term or diamagnetic current vanishes exactly for this model. For the vertices $v$ in Eqs.~\eqref{v0}--\eqref{v2}, the trace assumes the general form
\begin{align}
-\frac{i}{2}\mathrm{tr}[DvDv] =-\frac{i}{2} \int dt\, dt'\frac{d\Omega\, d\Omega'}{(2\pi)^2}e^{-i\Omega(t'-t)}e^{-i\Omega'(t-t')}\sum_{n,m,n,m'}\frac{i(nm|v(t)|n'm')i(n'm'|v(t')|nm)}{(\Omega-\xi_n +i\epsilon\, \mathrm{sgn}(\xi_n))(\Omega' - \xi_{n'}+i\epsilon\, \mathrm{sgn}(\xi_{n'}))}
\end{align}.
The denominator does not depend on the angular momenta $m$ and $m'$. Thus, the trace over $\{|m\ran, |m'\ran\}$ space can be computed separately. To this end, we decompose the vertices~\eqref{v0}--\eqref{v2} into 
\begin{subequations}
\begin{align}
A_0  & = \int \frac{d^3k}{(2\pi)^3}\widetilde{A}_0(k) e^{-i\omega t}e^{-|\ms k|^2}\ga^0(\ms k)e^{ib^{\dag}\bar{\ms k}}e^{ib\ms k},\\
\frac{-1}{2m_{\star}}\{\widehat{\Pi}_x, A_x\}  & =\int \frac{d^3k}{(2\pi)^3} \widetilde A_x(k)e^{-i\omega t}e^{-|\ms k|^2}  \ga^x(\ms k) e^{ib^{\dag}\bar{\ms k}}e^{ib\ms k},\\
\frac{-1}{2m_{\star}}\{\widehat{\Pi}_y, A_y\}  & =\int \frac{d^3k}{(2\pi)^3} \widetilde A_y(k)e^{-i\omega t}e^{-|\ms k|^2}  \ga^y(\ms k) e^{ib^{\dag}\bar{\ms k}}e^{ib\ms k}.
\end{align}
\end{subequations}
In this way, we trace out $\lan m |e^{ib^{\dag}\bar{\ms k}}e^{ib\ms k}|m'\ran\lan m'|e^{ib^{\dag}\bar{\ms k}'}e^{ib\ms k'}|m\ran$, obtaining a delta function $\frac{2\pi}{\ell^2}e^{|\ms k|^2}\delta^{(2)}(\mb k+\mb k')$. The time and frequency integrals for $(t, t')$ and $(\Omega, \Omega')$ can be performed with $\delta$ functions and residue calculus. The non-trivial summands remaining incorporate only the virtual processes between filled and empty Landau levels [$\mathrm{sgn}(\xi_n)\mathrm{sgn}(\xi_{n'})<0$]. Suppose the Fermi energy is pinned between the $N$th and the $(N+1)$th Landau levels as shown in Fig. \ref{fig3}. We then have
\begin{align}
\label{GF}-\frac{i}{2}\mathrm{tr}[DvDv] =-\frac{1}{2\pi\ell^2}\int \frac{d^3k}{(2\pi)^3}e^{-|\ms k|^2}\widetilde{A}_{\mu}(-k) \widetilde A_{\nu}(k)\sum_{n>N,n'\leq N}\bigg[\frac{\ga^{\mu}_{nn'}(-\ms k)\ga^{\nu}_{n'n}(\ms k)}{ \xi_{n'}-\xi_n+\omega+i\epsilon}+\frac{\ga^{\mu}_{n'n}(-\ms k)\ga^{\nu}_{nn'}(\ms k)}{\xi_{n'}-\xi_n-\omega+i\epsilon}\bigg],
\end{align}
with $\ga^{\mu}_{nn'} $ denoting the matrix element $(n|\ga^{\mu}|n')$. Equation~\eqref{GF} is an exact result of the trace to all orders in frequency and wavenumbers. The matrix elements $\ga^{\mu}_{nn'}(\ms k)$ can be written as linear combinations of the associated Laguerre polynomials $L^{\alpha}_n(|\ms k|^2)$. Some useful identities are documented in Appendix \ref{algebras}. The long-wavelength and low-frequency effective theory is derived upon expanding the $\Pi^{\mu\nu}(k)$ in small $\omega$ and $|\ms k|$. By comparing Eqs.~\eqref{GF} and~\eqref{Seff}, the time-ordered polarization tensor is identified as
\begin{align}\label{GFpolar}
 \Pi^{\mu\nu}(\omega, \mb k) = -\frac{e^{-|\ms k|^2}}{\pi\ell^2}\sum_{n>N,n'\leq N}\bigg[\frac{\ga^{\mu}_{nn'}(-\ms k)\ga^{\nu}_{n'n}(\ms k)}{ \xi_{n'}-\xi_n+\omega+i\epsilon}+\frac{\ga^{\mu}_{n'n}(-\ms k)\ga^{\nu}_{nn'}(\ms k)}{\xi_{n'}-\xi_n-\omega+i\epsilon}\bigg].
\end{align}
\end{widetext}
\subsection{Polarization tensors and transport coefficients}
Equation~\eqref{GFpolar} contains complete information about transport in our model of bilayer graphene. To avoid the ambiguity of doubly degenerate bands, we fill them simultaneously by taking $N>2$ and compute the polarization tensor to the specified limit or desired order in momentum. The transport observables can be extracted after properly modifying the epsilon prescription. For instance, with a homogeneous external electric field of form $\mb E = \mb E_{\omega}e^{-i\omega t}$, $\Pi^{00} = \Pi^{0i} = 0$ and the retarded $\Pi^{xy}_R$ can be computed exactly to all orders in frequency,
\begin{align}
\Pi^{xy}_R= \frac{i\omega}{2\pi} \frac{4N^3\omega_c^4 - 2N\omega^2\omega_c^2}{\omega_+^4 - 4N^2\omega_c^2(\omega_+^2-\omega_c^2)},
\end{align}
with $\omega_+ = \omega + i\epsilon.$

In the presence of a spatial fluctuation, at large scale, the effective action is organized by powers of $(\ell\p_i)$ and $\p_t/\omega_c$. By expanding and evaluating the sum to order $O(\p^3)$, the targeted parity-odd part of the Lagrangian is
\begin{align}
\mathscr L_{\rm eff}^{(\rm odd)} = -\frac{N}{4\pi} A\, dA &- \frac{3N^2-1}{4\pi}\frac{\ell^2}{2}\mb E\cdot\nabla B\notag\\
&-\frac{2N^2-1}{4\pi N}\frac{1}{2\omega_c^2}\epsilon^{ij}E_i\p_tE_j.
\end{align}

For nonrelativistic fermions, the level corresponds precisely to the number of filled Landau levels. The level here, $N$, is the index of the largest filled Landau level. As opposed to the bilayer graphene model, the number of filled Landau levels, and thus the interpretation of $N$, could be ambiguous due to the filled negative energy bands. This ambiguity is resolved by understanding the net contribution from the Fermi sea. By redoing the computation for $N= -1$, it is straightforward to confirm that negative energy bands contribute to $\frac{1}{4\pi}A\, dA$. Therefore, $N+1$ is understood as the number of filled bands with non-negative energies. 

We refer to the coefficient of the term $\frac{1}{2}(A_x\p_tA_y-A_y\p_tA_x)$ as the Hall conductivity $\sigma_H$ \footnote{This is not the most conventional definition, yet it is consistent when the extra minus sign from the electron charge is accounted for.}. To order $(q\ell)^2$ and $\omega^2$, the Hall conductivity is
\begin{align}
&\label{HallCon}\frac{\sigma_H }{\sigma_H(0)}= 1 + (q\ell)^2 \bigg(\frac{N^2+1}{4N}-N\bigg)+ \frac{\omega^2}{\omega_c^2}\frac{2N^2-1}{2N^2}
\end{align}
with $\sigma_H(0) = N/(2\pi).$ To make more sense out of this result, we can look at the static large $N$ limit, in which the hole band becomes insignificant and we should recover the physics of non-relativistic 2DEG. Taking $N\to\infty$, Eq.~\eqref{HallCon} can be organized into the form 
\begin{align*}
\frac{\sigma_H}{N/(2\pi)} = 1 + (q\ell)^2 \bigg( \frac{N^2/(8\pi\ell^2)}{N/(2\pi\ell^2)} - N\bigg).
\end{align*} 
Formally, it produces the Hoyos--Son relation \cite{PhysRevLett.108.066805, PhysRevB.94.125427} for 2DEG,
\begin{align}
\frac{\sigma_H}{\sigma_H(0)} = 1 + (q\ell)^2 \bigg[\frac{\eta_H}{n}-\frac{2\pi}{\nu}\frac{\ell^2}{\omega_c}B^2\epsilon''(B)\bigg],
\end{align}
if we identify the number density as $N/(2\pi\ell^2)$.

The Hoyos--Son result establishes a relation between the Hall viscosity $\eta_H(\omega)$ at long wavelengths, orbital magnetic susceptibility $-\frac{\p^2}{\p B^2}\epsilon(B)$, and the coefficient of Hall conductivity at order $(q\ell)^2$ based on the Galilean symmetry of the microscopic physics. In addition, the 2DEG result is physically understood as the ratio of $\eta_H$ to the charge density $n$. Starting from the model Eq.~\eqref{model}, we would not expect this relation to persist at finite $N$ because it lacks Galilean symmetry and the charge density depends on the regularization of the bottom of the Fermi sea. Nonetheless, motivated by observations at large $N$, we explore if a similar or approximate relation exists without any particular space-time symmetry. In particular, we wish to clarify the roles of Hall viscosity and orbital magnetic susceptibility. 

\section{Hall viscosity and orbital magnetic susceptibility of bilayer graphene}\label{VisSus}
To proceed with the last observation, our goal is to compute the Hall viscosity $\eta_H$ and the orbital magnetic susceptibility of the model Eq.~\eqref{model}. Within the framework of linear response, $\eta_H$ is given by the correlation function of the stress tensors $\lan \tau_{xx}\tau_{xy}\ran$, which can be expressed using the Kubo formula as \cite{PhysRevB.76.161305, PhysRevB.94.125427}
\begin{align}
\label{etaKubo}\eta_H(\omega) = \frac{1}{\pi\ell^2}\mathrm{Im}\sum_{a,b}\frac{(\tau_{xx})_{ba}(\tau_{xy})_{ab}}{\omega^2_+-(\xi_a-\xi_{b})^2},
\end{align}
where $a,b$ label bands above and below the Fermi energy. Defining the stress tensor nevertheless requires some caution for the following reason: The stress tensor can be defined as the response of the Hamiltonian with respect to the variation of the metric $g_{ij}$, but there is no obvious covariant way of coupling \eqref{model} to a general curved manifold. We circumvent this conceptual obstacle as follows: Instead of the projected low-energy Hamiltonian~\eqref{model}, we revisit the original model consisting of two Dirac spinors, where one consistently defines the action over a curved manifold \cite{PhysRevB.95.085151}, and hence the stress tensor $\tau_{ij}$. We derive the stress tensor in this manner and again project the components to the low-energy bands \cite{PhysRevB.101.035117}. The components of $\tau_{ij}$ are found to be 
\begin{subequations}
\begin{align}
\label{txx}& \tau_{xx} = -\frac{1}{m_{\star}}\pi_x^2 \sigma_x - \frac{1}{2m_{\star}}\sigma_y\{\pi_x, \pi_y\},\\
\label{tyy}& \tau_{yy} = \frac{1}{m_{\star}}\pi_y^2 \sigma_x - \frac{1}{2m_{\star}}\sigma_y \{\pi_x,\pi_y\},\\
\label{txy}& \tau_{xy} = -\frac{1}{4m_{\star}}\sigma_y \{\pi^{\dag},\pi\}.
\end{align}
\end{subequations}
In Appendix \ref{deriveT}, we present another distinct way to derive the above results, where a natural conjecture for the model on a general curved manifold is postulated and $\tau_{ij}$ is obtained by varying the Hamiltonian with respect to the vielbein field. 
A recent work \onlinecite{2019arXiv190909608I} derives the stress tensor by constructing strain generators and computing their commutators with the model Hamiltonian\cite{PhysRevB.86.245309}. This approach does not evade the difficulty discussed above. Taking the pseudospin degree of freedom as an example, it is not obvious that the pseudospin matrices generate physical rotations and should be included in the strain generators. 

Plugging $\tau_{xx}$ and $\tau_{xy}$ into Eq.~\eqref{etaKubo} yields the Hall viscosity to all orders in frequency. Particularly in the static limit $\omega \to 0$, we have 
\begin{align}
\label{etaH}\eta_H(\omega = 0) = \frac{1}{8\pi \ell^2}(N^2+1).
\end{align}

Caution is required to compute the orbital magnetic susceptibility $-\epsilon''(B)$, which is the negative of the second derivative of the energy density as a function of the background field. Suppose we naively evaluate the energy density per Landau level, $\epsilon(B)$:
\begin{subequations}
\begin{align}
\epsilon(B) = \frac{\omega_c}{2\pi\ell^2}\sum_{n = -\infty}^N \mathrm{sgn}(n)\sqrt{|n|(|n|-1)}.
\end{align}
After canceling the summands from $n = -N$ to $n =N$, the remaining sum is proportional to $\sum_{n=N+1}^{\infty}\sqrt{n(n-1)}$, which is severely divergent. To extract a finite result, we regularize the sum by introducing a natural cutoff $N_c$ obeying $\omega_c\sqrt{N_c(N_c-1)} = m_{\star}$, which constrains the validity of the low energy model. The regularized sum now reads 
\begin{align}
\epsilon(B) = -\frac{\omega_c}{2\pi\ell^2}\sum^{N_c}_{n=N+1}\sqrt{n(n-1)}.
\end{align}
\end{subequations}
The sum can be evaluated approximately with the Euler--Maclaurin formula. In a double expansion of large $N$ and small $\omega_c/m_{\star}$, the leading contribution is
\begin{align}
\label{chiB}-\epsilon''(B) \approx -\frac{N^2}{2\pi} + \frac{1}{16\pi}\ln \frac{N\omega_c}{m_{\star}}.
\end{align}
Since $N$ is bounded from above by $N_c\sim m_{\star}/\omega_c$, at large $N$ the logarithmic term is sub-leading. Consequently, both Eq.~\eqref{etaH} and Eq.~\eqref{chiB} coincide with the 2DEG results in the limit $N\to\infty$ \footnote{We note that the cutoff-dependent logarithm is not an artifact of the sharp cutoff, it also appears in other regularization schemes.}. In the rest of this paper, we exploit these observations and establish a concrete algebraic identity. 

\section{An algebraic relation from the Kubo formula}\label{KuboCurrent}
To better understand the relationship, we reexamine the Hall conductivity from the more conventional perspective of the Kubo formula. It is algebraically equivalent to the computation of the two-point functions $\Pi^{\mu\nu}$. Nonetheless, as we will show in a moment, it can make our speculation more transparent. The current operator in the first quantized form is given by the symmetrized velocity operator 
\begin{align}
\label{firstQuanCurrent}j^i(\mb r) =\frac{1}{2m_{\star}}\sum_k \{\widehat{\Pi}^i_k, \delta(\mb r-\mb r_k)\}.
\end{align}

Applying the Kubo formula for Hall conductivity to the current~\eqref{firstQuanCurrent}, we obtain 
\begin{align}
\label{KuboCon}\sigma_{H} = \mathrm{Im}\sum_{a,b}\frac{\{\widehat{\Pi}^x,e^{\frac{-\ms q\ell}{\sqrt 2}\pi^{\dag}}e^{\frac{\bar{\ms q}\ell}{\sqrt 2}\pi}\}_{ba}\{\widehat{\Pi}^y,e^{\frac{\ms q\ell}{\sqrt 2}\pi^{\dag}}e^{\frac{-\bar{\ms q}\ell}{\sqrt 2}\pi}\}_{ab}}{4m^2_{\star}\pi\ell^2e^{|\ms q|^2}[\omega^2_+ - (\xi_{a}-\xi_b)^2]}.
\end{align}
This formula can be quickly justified by considering $\frac{1}{i\omega}\Pi^{ij}$, which reduces to the conductivity tensor in the temporal gauge $A_0 = 0$. The above formula can be straightforwardly expanded in small $\ms q = (q\ell)$. To proceed, let us exploit rotational invariance and take $\mb q = (0,q)$. 
\begin{align*}
& \{\widehat{\Pi}^x, e^{\frac{-\mathsf q\ell}{\sqrt 2}\pi^{\dagger}}e^{\frac{\bar{\mathsf q}\ell}{\sqrt 2}\pi}\}\notag\\
\approx &\widehat{\Pi}^x (1+(q\ell)^2/4) - \frac{iq\tau_{xx}}{\omega_c} -\frac{q^2\ell^4}{4}\{\pi_x^2, \widehat{\Pi}^x\}.\\
& \{\widehat{\Pi}^y, e^{\frac{\mathsf q\ell}{\sqrt 2}\pi^{\dag}}e^{\frac{-\bar{\mathsf q}\ell}{\sqrt 2}\pi}\}\notag\\
\approx & \widehat{\Pi}^y(1+(q\ell)^2/4) + \frac{2iq\tau_{xy}}{\omega_c} + \frac{q\sigma_z\tau_{yy}}{\omega_c} -\frac{q^2\ell^2}{4}\{\pi_x^2, \widehat{\Pi}_y\}.
\end{align*}
To organize the products, we first observe the current operators map the $n$th Landau level to $(n\pm 1)$th, whereas the stress tensor operators only generate transitions between $n\to n,n\pm 2$. Their products thus have no finite summand and as a result, there is no term linear in $q$. Another non-trivial observation is at the level of the matrix elements $(\tau_{xx})_{ba}(\tau_{xy})_{ab} = (\tau_{xx})_{ba}(i\sigma_z\tau_{yy})_{ab}$ even though $\tau_{xy}\neq(i\sigma_z\tau_{yy})_{ab}$ in general. Consequently, expanding the numerator of~\eqref{KuboCon} to the second order in $(q\ell)$ yields the following identity,
\begin{align}\label{ConVisRelation}
&\sigma_H(\omega, q)\approx \sigma_H(\omega) + (q\ell)^2\bigg[\ell^2 \eta_H(\omega)\notag\\
& -\mathrm{Im}\sum_{a,b}\frac{(\widehat{\Pi}^x)_{ba}\{\pi_x^2,\widehat{\Pi}^y\}_{ab}+\{\pi^2_x,\widehat{\Pi}^x\}_{ba}(\widehat{\Pi}^y)_{ab}}{4\pi m^2_{\star}[\omega^2_+-(\xi_{a}-\xi_{b})^2]}\bigg],
\end{align}
which unambiguously identifies the role of Hall viscosity as part of the coefficient of $(q\ell)^2$. The sum in the second line of Eq.~\eqref{ConVisRelation} is not expressed directly with a physical observable at finite $\omega$. In the static limit, it sums to $-\frac{N^2}{2\pi} = -\epsilon''(B)-\frac{1}{16}\ln \frac{N}{N_c}$, and it reduces to Eq.~\eqref{main2} in the static limit $\omega \to 0$. At finite frequency, despite its lack of lucid physical interpretation, Eq.~\eqref{ConVisRelation} provides a decomposition that generates corrections to the Hoyos--Son relation in powers of $\omega^2$ and $N^{-1}$. 

\section{Discussion and conclusion}
The model Eq.~\eqref{model} is often considered a hybridization of Dirac and non-relativistic fermions, capturing the particle-hole structure of the former and the massive parabolic dispersion of the latter. The results derived in the main text entail in many ways that the features of non-relativistic fermions, or the restoration of Galilean symmetry, manifest asymptotically in the limit of a large filling factor. Explicit examples are the forms of Hall conductivity, Hall viscosity, orbital magnetic susceptibility, poles of transport coefficients, and the Hoyos--Son relation, although, in term of hydrodynamic relations \cite{doi:10.1142/S0217984989001400}, it is not yet clear in what sense the charge current and momentum density approach each other in the same limit. Established exact formulae can be utilized for interpolating between the asymmetric model and the Galilean symmetric paradigm. 

To move forward, this paper opens various directions. Empirical ones include the real-time effective theory at finite temperature using the Schwinger--Keldysh formalism \cite{PhysRevB.97.115123},  and interaction-generated transport properties owing to either an instantaneous Coulomb interaction or a mixed-dimensional Maxwell term \cite{PhysRevX.5.011040,PhysRevB.96.075127}. These developments will be critical in order to connect the single-particle toy model in the quantum Hall regime with experimental investigations of graphene materials \cite{Berdyugin162}, which are usually conducted in a hydrodynamic regime with strong disorder or interactions \cite{PhysRevB.100.035125, PhysRevB.100.115434}. Equally interesting are the inclusion of lattice effects in the generalized Hoyos--Son relation \cite{PhysRevB.98.245303}, generalization of the linear response theory to graphite multilayers as in Ref. \onlinecite{PhysRevB.101.155310}, and clarifying the distinction between dual descriptions of the $\nu = 1$ fractional quantum Hall state \cite{PhysRevLett.117.136802, PhysRevB.100.235150}. 

To conclude, we have determined the time-reversal odd electromagnetic response for the low-energy model of bilayer graphene to quadratic order in momentum and frequency, endowed a precise definition of the stress tensor to the low-energy projected model, and established a conductivity-viscosity relationship in the absence of obvious space-time symmetry. We investigated the limit in which the symmetry is restored and provided support from concrete computation at the operator level. In addition to the conclusions in the above, the effective action derived and the vielbein formulation introduced in this work can be further applied to the exploration of unknown facets of this model.  
\acknowledgements
The author thanks Yu-Ping Lin, M. Lapa and F Setiawan for comments on the manuscript. This work is supported by a Simons Investigator Grant from the Simons Foundation. 

\appendix

\section{Coherent state algebras}\label{algebras}
\subsection{Computing traces involving $b$ operators}
In the two-point function, the trace over angular momentum subspace can be performed separately, since the denominator of the propagator does not involve the angular momentum quantum number. The quantity we wish to compute is
\begin{align}
& \sum_{m,m'} \lan m|e^{ib^{\dag}\bar{\ms k}}e^{ib\ms k}|m'\ran\lan m'|e^{ib^{\dag}\bar{\ms k}'}e^{ib\ms k'}|m\ran\notag\\
= & \sum_{m} \lan m|e^{ib^{\dag}\bar{\ms k}}e^{ib\ms k}e^{ib^{\dag}\bar{\ms k}'}e^{ib\ms k'}|m\ran \notag.
\end{align}
From here we replace the sum over $m$ with an integral over all coherent states. 
\begin{align}
& \int d\mu(q)\lan q|e^{ib^{\dag}\bar{\ms k}}e^{ib\ms k}e^{ib^{\dag}\bar{\ms k}'}e^{ib\ms k'}|q\ran\notag\\
= & \int d\mu(q)\, e^{-|q|^2}\lan 0 |e^{\bar qb}e^{ib^{\dag}\bar{\ms k}}e^{ib\ms k}e^{ib^{\dag}\bar{\ms k}'}e^{ib\ms k'}e^{qb^{\dag}}|0\ran.
\end{align}
Since $[b,b^{\dag}]=1$, we can use the identity $e^Ae^B = e^Be^Ae^{[A,B]}$ and simplify the brackets to get
\begin{align}
\pi e^{|\ms k|^2}\delta^{(2)}(\ms k+\ms k') = \frac{2\pi}{\ell^2}e^{|\ms k|^2}\delta^{(2)}(\mb k+\mb k').
\end{align}

\subsection{Matrix elements $\ga^{\mu}_{nn'}(\ms k)$}
Here we document the matrix elements for the vertices used in the main text. The fundamental ingredient is $\lan n|e^{ia^{\dag}\ms k}e^{ia\bar{\ms k}}|n'\ran$. 
Let us evaluate 
\begin{align}
& \langle n |e^{ia^{\dagger}\mathsf k}e^{ia\bar{\mathsf k}}|n'\rangle = \sum_{s,s'}\langle n|\frac{(ia^{\dagger}\mathsf k)^s}{s!}\frac{(ia\bar{\mathsf k})^{s'}}{s'!}|n'\rangle \notag\\
= & \sum^n_s\sum^{n'}_{s'}\frac{(i\mathsf k)^s(i\bar{\mathsf k})^{s'}}{s!s'!(n-s)!}\delta_{n-s,n'-s'}\sqrt{n!n'!}
\end{align}
\begin{subequations}
The sum can be expressed in terms of generalized Laguerre polynomials $L^m_n(x)$. If $n<n'$, we have to evaluate $s'$ at $s' = s+n'-n$,
\begin{align}
 \lan n |e^{ia^{\dag}\ms k}e^{ia\bar{\ms k}}|n'\ran = (i\bar{\ms k})^{n'-n}\sqrt{\frac{n!}{n'!}}L^{n'-n}_n(|\ms k|^2).
\end{align}
On the other hand, if $n\geq n'$, we have to evaluate at $s = s' + n-n'$
\begin{align}
 \lan n |e^{ia^{\dag}\ms k}e^{ia\bar{\ms k}}|n'\ran  = (i\ms k)^{n-n'}\sqrt{\frac{n'!}{n!}}L^{n-n'}_{n'}(|\ms k|^2).
\end{align}
\end{subequations}
Together with the eigenstates~\eqref{zerobands} and~\eqref{otherbands}, it is then straightforward to compute $\ga^{\mu}_{nn'}(\ms k)$ and $\ga^{\mu}_{n'n}(\ms k) = [\ga^{\mu}_{nn'}(-\ms k)]^*$. 
\begin{widetext}
For definiteness, we consider the case $n>1$ and $n'<n$.
\paragraph{$\ga^0_{nn'}(\ms k)$}
For $n' = 0$ and $1$,
\begin{align}
& \ga_{n0}^0(\ms k) = \frac{1}{\sqrt{2(|n|!)}}(i\ms k)^{|n|}\\
& \ga_{n1}^0(\ms k) = \frac{1}{\sqrt{2(n!)}}(i\ms k)^{n-1}L_1^{n-1}(|\ms k|^2)
\end{align}
For $|n'|>1$, 
\begin{equation}
\ga_{nn'} =\left\{ \begin{array}{ll}\frac{\mathrm{sgn}(n')}{2}(i\ms k)^{n-|n'|}\sqrt{\frac{(|n'|-2)!}{(n-2)!}}L^{n-|n'|}_{|n'|-2}(|\ms k|^2)+\frac{1}{2}(i\ms k)^{n-|n'|}\sqrt{\frac{|n'|!}{n!}}L^{n-|n'|}_{|n'|}(|\ms k|^2), & n\geq |n'| \\ 
\frac{1}{2}(i\bar{\ms k})^{|n'|-n}\sqrt{\frac{n!}{|n'|!}}L^{|n'|-n}_n(|\ms k|^2)+\frac{\mathrm{sgn}(n')}{2}(i\bar{\ms k})^{|n'|-n}\sqrt{\frac{(n-2)!}{(|n'|-2)!}}L^{|n'|-n}_{n-2}(|\ms k|^2), & n<|n'|\end{array}\right. .
\end{equation}
\paragraph{$\ga^{i}_{nn'}(\ms k)$} It is convenient to define
\begin{align}
\ga^+(\ms k) = \frac{1}{2}(\sigma_x + i\sigma_y) \{a, e^{ia^{\dag}\ms k}e^{ia\bar{\ms k}}\}, \ga^-(\ms k) = [\ga^+(-\ms k)]^{\dag},
\end{align}
in terms of which,
\begin{subequations}
\begin{align}
&\ga^x(\ms k) = \frac{i}{\sqrt{2}\, m_{\star}\ell}[\ga^+(\ms k)-\ga^-(\ms k)],\\
&\ga^y(\ms k) = \frac{1}{\sqrt{2}\, m_{\star}\ell}[\ga^+(\ms k) + \ga^-(\ms k)].
\end{align}
\end{subequations}
For $n>1$,
\begin{equation}
\ga^+_{nn'}(\ms k) = \left\{ \begin{array}{ll} \frac{1}{2}(i\ms k)^{n-|n'|-1}\sqrt{\frac{|n'|!}{(n-2)!}}\bigg[L^{n-|n'|-1}_{|n'|}(|\ms k|^2) + L^{n-|n'|-1}_{|n'|-1}(|\ms k|^2)\bigg],& n\geq |n'|+1 \\ \frac{1}{2}(i\bar{\ms k})^{|n'|-n+1}\sqrt{\frac{(n-2)!}{|n'|!}}\bigg[(n-1)L^{|n'|-n+1}_{n-1}(|\ms k|^2) + |n'|L^{|n'|-n+1}_{n-2}(|\ms k|^2)\bigg], & n< |n'|+1\end{array} \right. .
\end{equation}
\begin{equation}
\ga^+_{n'n}(\ms k) = \left\{ \begin{array}{ll} \frac{\mathrm{sgn}(n')}{2}(i\ms k)^{|n'|-n-1}\sqrt{\frac{n!}{(|n'|-2)!}}\bigg[L^{|n'|-1-n}_n(|\ms k|^2) +L^{|n'|-n-1}_{n-1}(|\ms k|^2)\bigg], & |n'|\geq n+1\\ \frac{\mathrm{sgn}(n')}{2}(i\bar{\ms k})^{n-|n'|+1}\sqrt{\frac{(|n'|-2)!}{n!}}\bigg[ (|n'|-1)L^{n-|n'|+1}_{|n'|-1}(|\ms k|^2) +n L^{n-|n'|+1}_{|n'|-2}(|\ms k|^2)\bigg], & |n'|<n+1\end{array}\right. .
\end{equation}
Using these matrix elements, the generating functional~\eqref{GF} can be computed in a straightforward manner. 
\end{widetext}

\section{Another derivation of stress tensor} \label{deriveT} 
We derive here the stress tensor with a conjectured curved space generalization of model~\eqref{model} \footnote{The author thanks Dam Thanh Son for suggesting a derivation from this perspective.}. Let us consider a curved space endowed with the metric $g_{ij}(\mb x)$. A matrix field vielbein $e^a_j(\mb x)$ can be introduced via the relation $g_{ij} = \delta_{ab}e^a_ie^b_j$, where $a,b = 1,2$ are local SO(2) rotation indices \cite{Nakahara:2003nw}. The inverse field $E^i_a$ fulfills $E^i_ae_i^b = \delta_a^b$ and $g^{ij} = \delta^{ab}E_a^iE_b^j$. With the vielbeins, the spatial gradient operator $\p_a$ can be promoted to a curved manifold following $\p_a \to E^i_a\p_i$. Let us further denote $E_{\pm}^i = E^i_1\pm iE^i_2$. The natural generalization of Hamiltonian~\eqref{model} in the curved space reads 
\begin{align}
\label{curvedH} H =\frac{1}{2m_{\star}}\int d^2\mb x\, [(E_-^i\pi_i\psi^{\dag})(E^j_-\pi_j \chi)+\mathrm{h.c.}],
\end{align}
where $\psi$ and $\chi$ are the Grassmannian density of $\phi_A$ and $\phi_{\tilde B}$ and ``h.c." refers to the Hermitian conjugate. To derive the stress tensor relevant for viscosity computations, we turn on a slightly curved manifold with the metric $g_{ij} = \delta_{ij}+\delta g_{ij}$ and assume $\delta g_{ij}$ to be homogeneous. Under this circumstance, $\tau_{ij}$ assumes the following form 
\begin{align}
\tau_{ij} = \frac{1}{2}\left[e^a_i\frac{\p H}{\p E^{j}_a}+e^a_j\frac{\p H}{\p E^{i}_a}\right].
\end{align}
Applying this formula to~\eqref{curvedH}, we arrive at identical results, Eqs.~\eqref{txx}--\eqref{txy}.
\bibliography{citation}

\begin{thebibliography}{36}%
\makeatletter
\providecommand \@ifxundefined [1]{%
 \@ifx{#1\undefined}
}%
\providecommand \@ifnum [1]{%
 \ifnum #1\expandafter \@firstoftwo
 \else \expandafter \@secondoftwo
 \fi
}%
\providecommand \@ifx [1]{%
 \ifx #1\expandafter \@firstoftwo
 \else \expandafter \@secondoftwo
 \fi
}%
\providecommand \natexlab [1]{#1}%
\providecommand \enquote  [1]{``#1''}%
\providecommand \bibnamefont  [1]{#1}%
\providecommand \bibfnamefont [1]{#1}%
\providecommand \citenamefont [1]{#1}%
\providecommand \href@noop [0]{\@secondoftwo}%
\providecommand \href [0]{\begingroup \@sanitize@url \@href}%
\providecommand \@href[1]{\@@startlink{#1}\@@href}%
\providecommand \@@href[1]{\endgroup#1\@@endlink}%
\providecommand \@sanitize@url [0]{\catcode `\\12\catcode `\$12\catcode
  `\&12\catcode `\#12\catcode `\^12\catcode `\_12\catcode `\%12\relax}%
\providecommand \@@startlink[1]{}%
\providecommand \@@endlink[0]{}%
\providecommand \url  [0]{\begingroup\@sanitize@url \@url }%
\providecommand \@url [1]{\endgroup\@href {#1}{\urlprefix }}%
\providecommand \urlprefix  [0]{URL }%
\providecommand \Eprint [0]{\href }%
\providecommand \doibase [0]{https://doi.org/}%
\providecommand \selectlanguage [0]{\@gobble}%
\providecommand \bibinfo  [0]{\@secondoftwo}%
\providecommand \bibfield  [0]{\@secondoftwo}%
\providecommand \translation [1]{[#1]}%
\providecommand \BibitemOpen [0]{}%
\providecommand \bibitemStop [0]{}%
\providecommand \bibitemNoStop [0]{.\EOS\space}%
\providecommand \EOS [0]{\spacefactor3000\relax}%
\providecommand \BibitemShut  [1]{\csname bibitem#1\endcsname}%
\let\auto@bib@innerbib\@empty
\bibitem [{\citenamefont {Castro~Neto}\ \emph {et~al.}(2009)\citenamefont
  {Castro~Neto}, \citenamefont {Guinea}, \citenamefont {Peres}, \citenamefont
  {Novoselov},\ and\ \citenamefont {Geim}}]{RevModPhys.81.109}%
  \BibitemOpen
  \bibfield  {author} {\bibinfo {author} {\bibfnamefont {A.~H.}\ \bibnamefont
  {Castro~Neto}}, \bibinfo {author} {\bibfnamefont {F.}~\bibnamefont {Guinea}},
  \bibinfo {author} {\bibfnamefont {N.~M.~R.}\ \bibnamefont {Peres}}, \bibinfo
  {author} {\bibfnamefont {K.~S.}\ \bibnamefont {Novoselov}},\ and\ \bibinfo
  {author} {\bibfnamefont {A.~K.}\ \bibnamefont {Geim}},\ }\bibfield  {title}
  {\bibinfo {title} {The electronic properties of graphene},\ }\href
  {https://doi.org/10.1103/RevModPhys.81.109} {\bibfield  {journal} {\bibinfo
  {journal} {Rev. Mod. Phys.}\ }\textbf {\bibinfo {volume} {81}},\ \bibinfo
  {pages} {109} (\bibinfo {year} {2009})}\BibitemShut {NoStop}%
\bibitem [{\citenamefont {Gusynin}\ and\ \citenamefont
  {Sharapov}(2005)}]{PhysRevLett.95.146801}%
  \BibitemOpen
  \bibfield  {author} {\bibinfo {author} {\bibfnamefont {V.~P.}\ \bibnamefont
  {Gusynin}}\ and\ \bibinfo {author} {\bibfnamefont {S.~G.}\ \bibnamefont
  {Sharapov}},\ }\bibfield  {title} {\bibinfo {title} {Unconventional integer
  quantum hall effect in graphene},\ }\href
  {https://doi.org/10.1103/PhysRevLett.95.146801} {\bibfield  {journal}
  {\bibinfo  {journal} {Phys. Rev. Lett.}\ }\textbf {\bibinfo {volume} {95}},\
  \bibinfo {pages} {146801} (\bibinfo {year} {2005})}\BibitemShut {NoStop}%
\bibitem [{\citenamefont {{McCann}}\ and\ \citenamefont
  {{Koshino}}(2013)}]{2013RPPh76e6503M}%
  \BibitemOpen
  \bibfield  {author} {\bibinfo {author} {\bibfnamefont {E.}~\bibnamefont
  {{McCann}}}\ and\ \bibinfo {author} {\bibfnamefont {M.}~\bibnamefont
  {{Koshino}}},\ }\bibfield  {title} {\bibinfo {title} {{The electronic
  properties of bilayer graphene}},\ }\href
  {https://doi.org/10.1088/0034-4885/76/5/056503} {\bibfield  {journal}
  {\bibinfo  {journal} {Reports on Progress in Physics}\ }\textbf {\bibinfo
  {volume} {76}},\ \bibinfo {eid} {056503} (\bibinfo {year} {2013})},\ \Eprint
  {https://arxiv.org/abs/1205.6953} {arXiv:1205.6953 [cond-mat.mes-hall]}
  \BibitemShut {NoStop}%
\bibitem [{\citenamefont {Mross}\ \emph {et~al.}(2016)\citenamefont {Mross},
  \citenamefont {Alicea},\ and\ \citenamefont
  {Motrunich}}]{PhysRevLett.117.136802}%
  \BibitemOpen
  \bibfield  {author} {\bibinfo {author} {\bibfnamefont {D.~F.}\ \bibnamefont
  {Mross}}, \bibinfo {author} {\bibfnamefont {J.}~\bibnamefont {Alicea}},\ and\
  \bibinfo {author} {\bibfnamefont {O.~I.}\ \bibnamefont {Motrunich}},\
  }\bibfield  {title} {\bibinfo {title} {Bosonic analogue of dirac composite
  fermi liquid},\ }\href {https://doi.org/10.1103/PhysRevLett.117.136802}
  {\bibfield  {journal} {\bibinfo  {journal} {Phys. Rev. Lett.}\ }\textbf
  {\bibinfo {volume} {117}},\ \bibinfo {pages} {136802} (\bibinfo {year}
  {2016})}\BibitemShut {NoStop}%
\bibitem [{\citenamefont {Moldovan}\ \emph {et~al.}(2020)\citenamefont
  {Moldovan}, \citenamefont {Andelkovic},\ and\ \citenamefont
  {Peeters}}]{dean_moldovan_2020_4010216}%
  \BibitemOpen
  \bibfield  {author} {\bibinfo {author} {\bibfnamefont {D.}~\bibnamefont
  {Moldovan}}, \bibinfo {author} {\bibfnamefont {M.}~\bibnamefont
  {Andelkovic}},\ and\ \bibinfo {author} {\bibfnamefont {F.}~\bibnamefont
  {Peeters}},\ }\href {https://doi.org/10.5281/zenodo.4010216} {\bibinfo
  {title} {{pybinding v0.9.5: a Python package for tight- binding
  calculations}}} (\bibinfo {year} {2020})\BibitemShut {NoStop}%
\bibitem [{\citenamefont {McCann}\ and\ \citenamefont
  {Fal'ko}(2006)}]{PhysRevLett.96.086805}%
  \BibitemOpen
  \bibfield  {author} {\bibinfo {author} {\bibfnamefont {E.}~\bibnamefont
  {McCann}}\ and\ \bibinfo {author} {\bibfnamefont {V.~I.}\ \bibnamefont
  {Fal'ko}},\ }\bibfield  {title} {\bibinfo {title} {Landau-level degeneracy
  and quantum hall effect in a graphite bilayer},\ }\href
  {https://doi.org/10.1103/PhysRevLett.96.086805} {\bibfield  {journal}
  {\bibinfo  {journal} {Phys. Rev. Lett.}\ }\textbf {\bibinfo {volume} {96}},\
  \bibinfo {pages} {086805} (\bibinfo {year} {2006})}\BibitemShut {NoStop}%
\bibitem [{\citenamefont {Novoselov}\ \emph {et~al.}(2006)\citenamefont
  {Novoselov}, \citenamefont {McCann}, \citenamefont {Morozov}, \citenamefont
  {Fal'ko}, \citenamefont {Katsnelson}, \citenamefont {Zeitler}, \citenamefont
  {Jiang}, \citenamefont {Schedin},\ and\ \citenamefont
  {Geim}}]{Novoselov2006}%
  \BibitemOpen
  \bibfield  {author} {\bibinfo {author} {\bibfnamefont {K.~S.}\ \bibnamefont
  {Novoselov}}, \bibinfo {author} {\bibfnamefont {E.}~\bibnamefont {McCann}},
  \bibinfo {author} {\bibfnamefont {S.~V.}\ \bibnamefont {Morozov}}, \bibinfo
  {author} {\bibfnamefont {V.~I.}\ \bibnamefont {Fal'ko}}, \bibinfo {author}
  {\bibfnamefont {M.~I.}\ \bibnamefont {Katsnelson}}, \bibinfo {author}
  {\bibfnamefont {U.}~\bibnamefont {Zeitler}}, \bibinfo {author} {\bibfnamefont
  {D.}~\bibnamefont {Jiang}}, \bibinfo {author} {\bibfnamefont
  {F.}~\bibnamefont {Schedin}},\ and\ \bibinfo {author} {\bibfnamefont {A.~K.}\
  \bibnamefont {Geim}},\ }\bibfield  {title} {\bibinfo {title} {Unconventional
  quantum hall effect and berry's phase of 2p in bilayer graphene},\ }\href
  {https://doi.org/10.1038/nphys245} {\bibfield  {journal} {\bibinfo  {journal}
  {Nature Physics}\ }\textbf {\bibinfo {volume} {2}},\ \bibinfo {pages} {177}
  (\bibinfo {year} {2006})}\BibitemShut {NoStop}%
\bibitem [{\citenamefont {Papi\ifmmode~\acute{c}\else \'{c}\fi{}}\ and\
  \citenamefont {Abanin}(2014)}]{PhysRevLett.112.046602}%
  \BibitemOpen
  \bibfield  {author} {\bibinfo {author} {\bibfnamefont {Z.}~\bibnamefont
  {Papi\ifmmode~\acute{c}\else \'{c}\fi{}}}\ and\ \bibinfo {author}
  {\bibfnamefont {D.~A.}\ \bibnamefont {Abanin}},\ }\bibfield  {title}
  {\bibinfo {title} {Topological phases in the zeroth landau level of bilayer
  graphene},\ }\href {https://doi.org/10.1103/PhysRevLett.112.046602}
  {\bibfield  {journal} {\bibinfo  {journal} {Phys. Rev. Lett.}\ }\textbf
  {\bibinfo {volume} {112}},\ \bibinfo {pages} {046602} (\bibinfo {year}
  {2014})}\BibitemShut {NoStop}%
\bibitem [{\citenamefont {Misumi}\ and\ \citenamefont
  {Shizuya}(2008)}]{PhysRevB.77.195423}%
  \BibitemOpen
  \bibfield  {author} {\bibinfo {author} {\bibfnamefont {T.}~\bibnamefont
  {Misumi}}\ and\ \bibinfo {author} {\bibfnamefont {K.}~\bibnamefont
  {Shizuya}},\ }\bibfield  {title} {\bibinfo {title} {Electromagnetic response
  and pseudo-zero-mode landau levels of bilayer graphene in a magnetic field},\
  }\href {https://doi.org/10.1103/PhysRevB.77.195423} {\bibfield  {journal}
  {\bibinfo  {journal} {Phys. Rev. B}\ }\textbf {\bibinfo {volume} {77}},\
  \bibinfo {pages} {195423} (\bibinfo {year} {2008})}\BibitemShut {NoStop}%
\bibitem [{\citenamefont {Shizuya}(2009)}]{PhysRevB.79.165402}%
  \BibitemOpen
  \bibfield  {author} {\bibinfo {author} {\bibfnamefont {K.}~\bibnamefont
  {Shizuya}},\ }\bibfield  {title} {\bibinfo {title} {Pseudo-zero-mode landau
  levels and collective excitations in bilayer graphene},\ }\href
  {https://doi.org/10.1103/PhysRevB.79.165402} {\bibfield  {journal} {\bibinfo
  {journal} {Phys. Rev. B}\ }\textbf {\bibinfo {volume} {79}},\ \bibinfo
  {pages} {165402} (\bibinfo {year} {2009})}\BibitemShut {NoStop}%
\bibitem [{\citenamefont {{Imran}}(2019)}]{2019arXiv190909608I}%
  \BibitemOpen
  \bibfield  {author} {\bibinfo {author} {\bibfnamefont {M.}~\bibnamefont
  {{Imran}}},\ }\bibfield  {title} {\bibinfo {title} {{Quantizing momentum
  transport in bilayer graphene}},\ }\href@noop {} {\bibfield  {journal}
  {\bibinfo  {journal} {arXiv e-prints}\ ,\ \bibinfo {eid} {arXiv:1909.09608}}
  (\bibinfo {year} {2019})},\ \Eprint {https://arxiv.org/abs/1909.09608}
  {arXiv:1909.09608 [cond-mat.mes-hall]} \BibitemShut {NoStop}%
\bibitem [{\citenamefont {Hoyos}\ and\ \citenamefont
  {Son}(2012)}]{PhysRevLett.108.066805}%
  \BibitemOpen
  \bibfield  {author} {\bibinfo {author} {\bibfnamefont {C.}~\bibnamefont
  {Hoyos}}\ and\ \bibinfo {author} {\bibfnamefont {D.~T.}\ \bibnamefont
  {Son}},\ }\bibfield  {title} {\bibinfo {title} {Hall viscosity and
  electromagnetic response},\ }\href
  {https://doi.org/10.1103/PhysRevLett.108.066805} {\bibfield  {journal}
  {\bibinfo  {journal} {Phys. Rev. Lett.}\ }\textbf {\bibinfo {volume} {108}},\
  \bibinfo {pages} {066805} (\bibinfo {year} {2012})}\BibitemShut {NoStop}%
\bibitem [{\citenamefont {Bradlyn}\ \emph {et~al.}(2012)\citenamefont
  {Bradlyn}, \citenamefont {Goldstein},\ and\ \citenamefont
  {Read}}]{PhysRevB.86.245309}%
  \BibitemOpen
  \bibfield  {author} {\bibinfo {author} {\bibfnamefont {B.}~\bibnamefont
  {Bradlyn}}, \bibinfo {author} {\bibfnamefont {M.}~\bibnamefont {Goldstein}},\
  and\ \bibinfo {author} {\bibfnamefont {N.}~\bibnamefont {Read}},\ }\bibfield
  {title} {\bibinfo {title} {Kubo formulas for viscosity: Hall viscosity, ward
  identities, and the relation with conductivity},\ }\href
  {https://doi.org/10.1103/PhysRevB.86.245309} {\bibfield  {journal} {\bibinfo
  {journal} {Phys. Rev. B}\ }\textbf {\bibinfo {volume} {86}},\ \bibinfo
  {pages} {245309} (\bibinfo {year} {2012})}\BibitemShut {NoStop}%
\bibitem [{\citenamefont {Rao}\ and\ \citenamefont
  {Bradlyn}(2020)}]{PhysRevX.10.021005}%
  \BibitemOpen
  \bibfield  {author} {\bibinfo {author} {\bibfnamefont {P.}~\bibnamefont
  {Rao}}\ and\ \bibinfo {author} {\bibfnamefont {B.}~\bibnamefont {Bradlyn}},\
  }\bibfield  {title} {\bibinfo {title} {Hall viscosity in quantum systems with
  discrete symmetry: Point group and lattice anisotropy},\ }\href
  {https://doi.org/10.1103/PhysRevX.10.021005} {\bibfield  {journal} {\bibinfo
  {journal} {Phys. Rev. X}\ }\textbf {\bibinfo {volume} {10}},\ \bibinfo
  {pages} {021005} (\bibinfo {year} {2020})}\BibitemShut {NoStop}%
\bibitem [{\citenamefont {Sherafati}\ \emph {et~al.}(2016)\citenamefont
  {Sherafati}, \citenamefont {Principi},\ and\ \citenamefont
  {Vignale}}]{PhysRevB.94.125427}%
  \BibitemOpen
  \bibfield  {author} {\bibinfo {author} {\bibfnamefont {M.}~\bibnamefont
  {Sherafati}}, \bibinfo {author} {\bibfnamefont {A.}~\bibnamefont
  {Principi}},\ and\ \bibinfo {author} {\bibfnamefont {G.}~\bibnamefont
  {Vignale}},\ }\bibfield  {title} {\bibinfo {title} {Hall viscosity and
  electromagnetic response of electrons in graphene},\ }\href
  {https://doi.org/10.1103/PhysRevB.94.125427} {\bibfield  {journal} {\bibinfo
  {journal} {Phys. Rev. B}\ }\textbf {\bibinfo {volume} {94}},\ \bibinfo
  {pages} {125427} (\bibinfo {year} {2016})}\BibitemShut {NoStop}%
\bibitem [{\citenamefont {Briskot}\ \emph {et~al.}(2015)\citenamefont
  {Briskot}, \citenamefont {Sch\"utt}, \citenamefont {Gornyi}, \citenamefont
  {Titov}, \citenamefont {Narozhny},\ and\ \citenamefont
  {Mirlin}}]{PhysRevB.92.115426}%
  \BibitemOpen
  \bibfield  {author} {\bibinfo {author} {\bibfnamefont {U.}~\bibnamefont
  {Briskot}}, \bibinfo {author} {\bibfnamefont {M.}~\bibnamefont {Sch\"utt}},
  \bibinfo {author} {\bibfnamefont {I.~V.}\ \bibnamefont {Gornyi}}, \bibinfo
  {author} {\bibfnamefont {M.}~\bibnamefont {Titov}}, \bibinfo {author}
  {\bibfnamefont {B.~N.}\ \bibnamefont {Narozhny}},\ and\ \bibinfo {author}
  {\bibfnamefont {A.~D.}\ \bibnamefont {Mirlin}},\ }\bibfield  {title}
  {\bibinfo {title} {Collision-dominated nonlinear hydrodynamics in graphene},\
  }\href {https://doi.org/10.1103/PhysRevB.92.115426} {\bibfield  {journal}
  {\bibinfo  {journal} {Phys. Rev. B}\ }\textbf {\bibinfo {volume} {92}},\
  \bibinfo {pages} {115426} (\bibinfo {year} {2015})}\BibitemShut {NoStop}%
\bibitem [{\citenamefont {Katsnelson}\ \emph {et~al.}(2013)\citenamefont
  {Katsnelson}, \citenamefont {Volovik},\ and\ \citenamefont
  {Zubkov}}]{KATSNELSON2013160}%
  \BibitemOpen
  \bibfield  {author} {\bibinfo {author} {\bibfnamefont {M.}~\bibnamefont
  {Katsnelson}}, \bibinfo {author} {\bibfnamefont {G.}~\bibnamefont
  {Volovik}},\ and\ \bibinfo {author} {\bibfnamefont {M.}~\bibnamefont
  {Zubkov}},\ }\bibfield  {title} {\bibinfo {title} {Euler–heisenberg
  effective action and magnetoelectric effect in multilayer graphene},\ }\href
  {https://doi.org/https://doi.org/10.1016/j.aop.2012.12.010} {\bibfield
  {journal} {\bibinfo  {journal} {Annals of Physics}\ }\textbf {\bibinfo
  {volume} {331}},\ \bibinfo {pages} {160 } (\bibinfo {year}
  {2013})}\BibitemShut {NoStop}%
\bibitem [{Note1()}]{Note1}%
  \BibitemOpen
  \bibinfo {note} {We use the notation $|\cdot )$ to denote the doublet wave
  functions.}\BibitemShut {Stop}%
\bibitem [{\citenamefont {Jain}(2007)}]{jain_2007}%
  \BibitemOpen
  \bibfield  {author} {\bibinfo {author} {\bibfnamefont {J.~K.}\ \bibnamefont
  {Jain}},\ }\href {https://doi.org/10.1017/CBO9780511607561} {\emph {\bibinfo
  {title} {Composite Fermions}}}\ (\bibinfo  {publisher} {Cambridge University
  Press},\ \bibinfo {year} {2007})\BibitemShut {NoStop}%
\bibitem [{Note2()}]{Note2}%
  \BibitemOpen
  \bibinfo {note} {This is not the most conventional definition, yet it is
  consistent when the extra minus sign from the electron charge is accounted
  for.}\BibitemShut {Stop}%
\bibitem [{\citenamefont {Tokatly}\ and\ \citenamefont
  {Vignale}(2007)}]{PhysRevB.76.161305}%
  \BibitemOpen
  \bibfield  {author} {\bibinfo {author} {\bibfnamefont {I.~V.}\ \bibnamefont
  {Tokatly}}\ and\ \bibinfo {author} {\bibfnamefont {G.}~\bibnamefont
  {Vignale}},\ }\bibfield  {title} {\bibinfo {title} {Lorentz shear modulus of
  a two-dimensional electron gas at high magnetic field},\ }\href
  {https://doi.org/10.1103/PhysRevB.76.161305} {\bibfield  {journal} {\bibinfo
  {journal} {Phys. Rev. B}\ }\textbf {\bibinfo {volume} {76}},\ \bibinfo
  {pages} {161305} (\bibinfo {year} {2007})}\BibitemShut {NoStop}%
\bibitem [{\citenamefont {Nguyen}\ and\ \citenamefont
  {Gromov}(2017)}]{PhysRevB.95.085151}%
  \BibitemOpen
  \bibfield  {author} {\bibinfo {author} {\bibfnamefont {D.~X.}\ \bibnamefont
  {Nguyen}}\ and\ \bibinfo {author} {\bibfnamefont {A.}~\bibnamefont
  {Gromov}},\ }\bibfield  {title} {\bibinfo {title} {Exact electromagnetic
  response of landau level electrons},\ }\href
  {https://doi.org/10.1103/PhysRevB.95.085151} {\bibfield  {journal} {\bibinfo
  {journal} {Phys. Rev. B}\ }\textbf {\bibinfo {volume} {95}},\ \bibinfo
  {pages} {085151} (\bibinfo {year} {2017})}\BibitemShut {NoStop}%
\bibitem [{\citenamefont {Nguyen}\ \emph {et~al.}(2020)\citenamefont {Nguyen},
  \citenamefont {Wagner},\ and\ \citenamefont {Simon}}]{PhysRevB.101.035117}%
  \BibitemOpen
  \bibfield  {author} {\bibinfo {author} {\bibfnamefont {D.~X.}\ \bibnamefont
  {Nguyen}}, \bibinfo {author} {\bibfnamefont {G.}~\bibnamefont {Wagner}},\
  and\ \bibinfo {author} {\bibfnamefont {S.~H.}\ \bibnamefont {Simon}},\
  }\bibfield  {title} {\bibinfo {title} {Quantum boltzmann equation for bilayer
  graphene},\ }\href {https://doi.org/10.1103/PhysRevB.101.035117} {\bibfield
  {journal} {\bibinfo  {journal} {Phys. Rev. B}\ }\textbf {\bibinfo {volume}
  {101}},\ \bibinfo {pages} {035117} (\bibinfo {year} {2020})}\BibitemShut
  {NoStop}%
\bibitem [{Note3()}]{Note3}%
  \BibitemOpen
  \bibinfo {note} {We note that the cutoff-dependent logarithm is not an
  artifact of the sharp cutoff, it also appears in other regularization
  schemes.}\BibitemShut {Stop}%
\bibitem [{\citenamefont {Greiter}\ \emph {et~al.}(1989)\citenamefont
  {Greiter}, \citenamefont {Wilczek},\ and\ \citenamefont
  {Witten}}]{doi:10.1142/S0217984989001400}%
  \BibitemOpen
  \bibfield  {author} {\bibinfo {author} {\bibfnamefont {M.}~\bibnamefont
  {Greiter}}, \bibinfo {author} {\bibfnamefont {F.}~\bibnamefont {Wilczek}},\
  and\ \bibinfo {author} {\bibfnamefont {E.}~\bibnamefont {Witten}},\
  }\bibfield  {title} {\bibinfo {title} {Hydrodynamic relations in
  superconductivity},\ }\href {https://doi.org/10.1142/S0217984989001400}
  {\bibfield  {journal} {\bibinfo  {journal} {Modern Physics Letters B}\
  }\textbf {\bibinfo {volume} {03}},\ \bibinfo {pages} {903} (\bibinfo {year}
  {1989})},\ \Eprint
  {https://arxiv.org/abs/https://doi.org/10.1142/S0217984989001400}
  {https://doi.org/10.1142/S0217984989001400} \BibitemShut {NoStop}%
\bibitem [{\citenamefont {Fr\"a\ss{}dorf}(2018)}]{PhysRevB.97.115123}%
  \BibitemOpen
  \bibfield  {author} {\bibinfo {author} {\bibfnamefont {C.}~\bibnamefont
  {Fr\"a\ss{}dorf}},\ }\bibfield  {title} {\bibinfo {title} {Abelian
  chern-simons theory for the fractional quantum hall effect in graphene},\
  }\href {https://doi.org/10.1103/PhysRevB.97.115123} {\bibfield  {journal}
  {\bibinfo  {journal} {Phys. Rev. B}\ }\textbf {\bibinfo {volume} {97}},\
  \bibinfo {pages} {115123} (\bibinfo {year} {2018})}\BibitemShut {NoStop}%
\bibitem [{\citenamefont {Marino}\ \emph {et~al.}(2015)\citenamefont {Marino},
  \citenamefont {Nascimento}, \citenamefont {Alves},\ and\ \citenamefont
  {Smith}}]{PhysRevX.5.011040}%
  \BibitemOpen
  \bibfield  {author} {\bibinfo {author} {\bibfnamefont {E.~C.}\ \bibnamefont
  {Marino}}, \bibinfo {author} {\bibfnamefont {L.~O.}\ \bibnamefont
  {Nascimento}}, \bibinfo {author} {\bibfnamefont {V.~S.}\ \bibnamefont
  {Alves}},\ and\ \bibinfo {author} {\bibfnamefont {C.~M.}\ \bibnamefont
  {Smith}},\ }\bibfield  {title} {\bibinfo {title} {Interaction induced quantum
  valley hall effect in graphene},\ }\href
  {https://doi.org/10.1103/PhysRevX.5.011040} {\bibfield  {journal} {\bibinfo
  {journal} {Phys. Rev. X}\ }\textbf {\bibinfo {volume} {5}},\ \bibinfo {pages}
  {011040} (\bibinfo {year} {2015})}\BibitemShut {NoStop}%
\bibitem [{\citenamefont {Hsiao}\ and\ \citenamefont
  {Son}(2017)}]{PhysRevB.96.075127}%
  \BibitemOpen
  \bibfield  {author} {\bibinfo {author} {\bibfnamefont {W.-H.}\ \bibnamefont
  {Hsiao}}\ and\ \bibinfo {author} {\bibfnamefont {D.~T.}\ \bibnamefont
  {Son}},\ }\bibfield  {title} {\bibinfo {title} {Duality and universal
  transport in mixed-dimension electrodynamics},\ }\href
  {https://doi.org/10.1103/PhysRevB.96.075127} {\bibfield  {journal} {\bibinfo
  {journal} {Phys. Rev. B}\ }\textbf {\bibinfo {volume} {96}},\ \bibinfo
  {pages} {075127} (\bibinfo {year} {2017})}\BibitemShut {NoStop}%
\bibitem [{\citenamefont {Berdyugin}\ \emph {et~al.}(2019)\citenamefont
  {Berdyugin}, \citenamefont {Xu}, \citenamefont {Pellegrino}, \citenamefont
  {Krishna~Kumar}, \citenamefont {Principi}, \citenamefont {Torre},
  \citenamefont {Ben~Shalom}, \citenamefont {Taniguchi}, \citenamefont
  {Watanabe}, \citenamefont {Grigorieva}, \citenamefont {Polini}, \citenamefont
  {Geim},\ and\ \citenamefont {Bandurin}}]{Berdyugin162}%
  \BibitemOpen
  \bibfield  {author} {\bibinfo {author} {\bibfnamefont {A.~I.}\ \bibnamefont
  {Berdyugin}}, \bibinfo {author} {\bibfnamefont {S.~G.}\ \bibnamefont {Xu}},
  \bibinfo {author} {\bibfnamefont {F.~M.~D.}\ \bibnamefont {Pellegrino}},
  \bibinfo {author} {\bibfnamefont {R.}~\bibnamefont {Krishna~Kumar}}, \bibinfo
  {author} {\bibfnamefont {A.}~\bibnamefont {Principi}}, \bibinfo {author}
  {\bibfnamefont {I.}~\bibnamefont {Torre}}, \bibinfo {author} {\bibfnamefont
  {M.}~\bibnamefont {Ben~Shalom}}, \bibinfo {author} {\bibfnamefont
  {T.}~\bibnamefont {Taniguchi}}, \bibinfo {author} {\bibfnamefont
  {K.}~\bibnamefont {Watanabe}}, \bibinfo {author} {\bibfnamefont {I.~V.}\
  \bibnamefont {Grigorieva}}, \bibinfo {author} {\bibfnamefont
  {M.}~\bibnamefont {Polini}}, \bibinfo {author} {\bibfnamefont {A.~K.}\
  \bibnamefont {Geim}},\ and\ \bibinfo {author} {\bibfnamefont {D.~A.}\
  \bibnamefont {Bandurin}},\ }\bibfield  {title} {\bibinfo {title} {Measuring
  hall viscosity of graphene{\textquoteright}s electron fluid},\ }\href
  {https://doi.org/10.1126/science.aau0685} {\bibfield  {journal} {\bibinfo
  {journal} {Science}\ }\textbf {\bibinfo {volume} {364}},\ \bibinfo {pages}
  {162} (\bibinfo {year} {2019})},\ \Eprint
  {https://arxiv.org/abs/https://science.sciencemag.org/content/364/6436/162.full.pdf}
  {https://science.sciencemag.org/content/364/6436/162.full.pdf} \BibitemShut
  {NoStop}%
\bibitem [{\citenamefont {Narozhny}\ and\ \citenamefont
  {Sch\"utt}(2019)}]{PhysRevB.100.035125}%
  \BibitemOpen
  \bibfield  {author} {\bibinfo {author} {\bibfnamefont {B.~N.}\ \bibnamefont
  {Narozhny}}\ and\ \bibinfo {author} {\bibfnamefont {M.}~\bibnamefont
  {Sch\"utt}},\ }\bibfield  {title} {\bibinfo {title} {Magnetohydrodynamics in
  graphene: Shear and hall viscosities},\ }\href
  {https://doi.org/10.1103/PhysRevB.100.035125} {\bibfield  {journal} {\bibinfo
   {journal} {Phys. Rev. B}\ }\textbf {\bibinfo {volume} {100}},\ \bibinfo
  {pages} {035125} (\bibinfo {year} {2019})}\BibitemShut {NoStop}%
\bibitem [{\citenamefont {Narozhny}(2019)}]{PhysRevB.100.115434}%
  \BibitemOpen
  \bibfield  {author} {\bibinfo {author} {\bibfnamefont {B.~N.}\ \bibnamefont
  {Narozhny}},\ }\bibfield  {title} {\bibinfo {title} {Optical conductivity in
  graphene: Hydrodynamic regime},\ }\href
  {https://doi.org/10.1103/PhysRevB.100.115434} {\bibfield  {journal} {\bibinfo
   {journal} {Phys. Rev. B}\ }\textbf {\bibinfo {volume} {100}},\ \bibinfo
  {pages} {115434} (\bibinfo {year} {2019})}\BibitemShut {NoStop}%
\bibitem [{\citenamefont {Harper}\ \emph {et~al.}(2018)\citenamefont {Harper},
  \citenamefont {Bauer}, \citenamefont {Jackson},\ and\ \citenamefont
  {Roy}}]{PhysRevB.98.245303}%
  \BibitemOpen
  \bibfield  {author} {\bibinfo {author} {\bibfnamefont {F.}~\bibnamefont
  {Harper}}, \bibinfo {author} {\bibfnamefont {D.}~\bibnamefont {Bauer}},
  \bibinfo {author} {\bibfnamefont {T.~S.}\ \bibnamefont {Jackson}},\ and\
  \bibinfo {author} {\bibfnamefont {R.}~\bibnamefont {Roy}},\ }\bibfield
  {title} {\bibinfo {title} {Finite-wave-vector electromagnetic response in
  lattice quantum hall systems},\ }\href
  {https://doi.org/10.1103/PhysRevB.98.245303} {\bibfield  {journal} {\bibinfo
  {journal} {Phys. Rev. B}\ }\textbf {\bibinfo {volume} {98}},\ \bibinfo
  {pages} {245303} (\bibinfo {year} {2018})}\BibitemShut {NoStop}%
\bibitem [{\citenamefont {Hsiao}(2020)}]{PhysRevB.101.155310}%
  \BibitemOpen
  \bibfield  {author} {\bibinfo {author} {\bibfnamefont {W.-H.}\ \bibnamefont
  {Hsiao}},\ }\bibfield  {title} {\bibinfo {title} {Landau quantization of
  multilayer graphene on a haldane sphere},\ }\href
  {https://doi.org/10.1103/PhysRevB.101.155310} {\bibfield  {journal} {\bibinfo
   {journal} {Phys. Rev. B}\ }\textbf {\bibinfo {volume} {101}},\ \bibinfo
  {pages} {155310} (\bibinfo {year} {2020})}\BibitemShut {NoStop}%
\bibitem [{\citenamefont {Hsiao}\ and\ \citenamefont
  {Son}(2019)}]{PhysRevB.100.235150}%
  \BibitemOpen
  \bibfield  {author} {\bibinfo {author} {\bibfnamefont {W.-H.}\ \bibnamefont
  {Hsiao}}\ and\ \bibinfo {author} {\bibfnamefont {D.~T.}\ \bibnamefont
  {Son}},\ }\bibfield  {title} {\bibinfo {title} {Self-dual
  $\ensuremath{\nu}=1$ bosonic quantum hall state in mixed-dimensional qed},\
  }\href {https://doi.org/10.1103/PhysRevB.100.235150} {\bibfield  {journal}
  {\bibinfo  {journal} {Phys. Rev. B}\ }\textbf {\bibinfo {volume} {100}},\
  \bibinfo {pages} {235150} (\bibinfo {year} {2019})}\BibitemShut {NoStop}%
\bibitem [{Note4()}]{Note4}%
  \BibitemOpen
  \bibinfo {note} {The author thanks Dam Thanh Son for suggesting a derivation
  from this perspective.}\BibitemShut {Stop}%
\bibitem [{\citenamefont {Nakahara}(2003)}]{Nakahara:2003nw}%
  \BibitemOpen
  \bibfield  {author} {\bibinfo {author} {\bibfnamefont {M.}~\bibnamefont
  {Nakahara}},\ }\href {https://books.google.com/books?id=cH-XQB0Ex5wC} {\emph
  {\bibinfo {title} {Geometry, Topology and Physics, Second Edition}}},\
  Graduate student series in physics\ (\bibinfo  {publisher} {Taylor \&
  Francis},\ \bibinfo {year} {2003})\BibitemShut {NoStop}%
\end{thebibliography}%
\end{document}